\documentclass[journal=jacsat,manuscript=article]{achemso}

\usepackage{array}
\usepackage{listing}
\usepackage{lmodern}
\usepackage{mathpazo}
\usepackage{microtype}

\author{Axel M. Eriksson}
\affiliation{Chalmers University of Technology, 41296 G\"oteborg, Sweden}
\email{marer@chalmers.se}
\author{Marina V. Voinova}
\affiliation{Chalmers University of Technology, 41296 G\"oteborg, Sweden}
\author{Leonid Y. Gorelik}

\affiliation{Chalmers University of Technology, 41296 G\"oteborg, Sweden}

\title{Nonresonant high frequency excitation of mechanical vibrations in graphene based nanoresonator}

\abbreviations{IR,NMR,UV}

\begin{document}

\begin{abstract}
We theoretically analyse the dynamics of a suspended graphene membrane which is in tunnel contact with grounded metallic electrodes and subjected to ac-electrostatic potential induced by a gate electrode.
It is shown that for such system the retardation effects in the electronic subsystem
generate an effective pumping for the relatively slow mechanical vibrations if the
driving frequency exceeds the inverse charge relaxation time. Under this condition there is a critical value of the driving voltage amplitude above which the pumping overcomes the intrinsic damping of the mechanical resonator leading to a mechanical instability. This nonresonant instability is saturated by nonlinear damping and the system exhibits self-sustained oscillations of relatively large amplitude.
\end{abstract}


Rapid progress in carbon nanostructures manufacturing stimulated new experimental and theoretical efforts in studying their unique optical, electrical, and mechanical properties\cite{Novoselov,Terrones2010}. In particular, the very high stiffness and low density of graphene make it an ideal material for the construction of nanoelectromechanical resonators. These graphene features are of great interest both for the fundamental studies of mechanics at the nanoscale level and a variety of applications, including force, position and mass sensing \cite{Jablan2013,Scott,Garcia,Chen2009,Hod}. In particular, it was demonstrated \cite{Chen2013} that the graphene-based nanomechanical resonator can be employed as an active element for frequency-modulated signal generation and efficient audio signal transmission. Operation of most of the nanomechanical devices is based on the excitation of mechanical vibrations by an external periodic force, of electrostatic or optic origin, with a frequency comparable with the vibrational frequency of the mechanical resonator\cite{Chen2013,Poot2010,Xu2010,Meerwaldt2012,Lassagne2009,Unterreithmeier2009}. At the same time, it was  shown \cite{Gorelik,Isacsson} that in certain nanoelectromechanical systems self-sustained  mechanical oscillations with relatively large amplitude may also be actuated by  ''shuttle instability''. In the shuttle structures described in \cite{Gorelik,Isacsson}, the instability was found to occur at driving frequencies which are much smaller compared with the eigenfrequency of the mechanical subsystem. In the present work we are seeking to answer the question if it is possible to achieve a regime of self-sustained oscillations in a graphene-based nanoresonator by using an electromechanical instability effect caused by a nonresonant driving field.
In the paper we demonstrate that such a possibility really exist. However, in the contrast to the shuttle instability, the electromechanical instability in the graphene-based resonators similar to those considered in the publication \cite{Scott,Garcia,Chen2009,Hod,Chen2013} occurs when the driving frequency is much greater than the eigenfrequency of the mechanical subsystem.

\begin{figure}\centering
\includegraphics[scale=0.5]{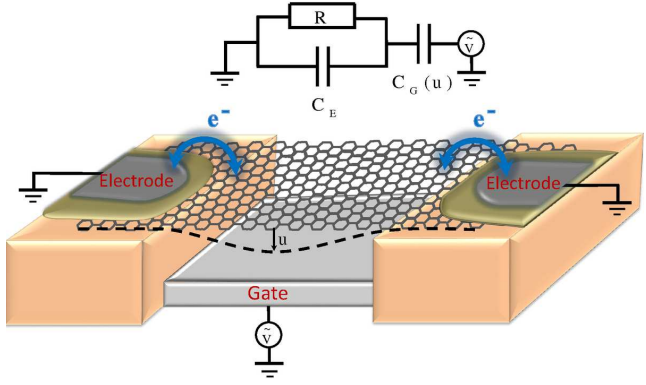}
\caption{A sketch of the graphene membrane resonator suspended over the trench and separated from the grounded metallic electrodes by an oxide layer. A side-gate subjected to an AC voltage induces an electrostatic potential on the graphene sheet. It is assumed that the potential is dependent on the membrane deflection  $u$. The schematic RC-circuit is shown above.}
\label{fig1}
\end{figure}

The system under consideration is shown in Figure \ref{fig1}. A doped graphene membrane is suspended over the trench and in contact with the  grounded metallic electrodes through the oxide layer.  An oscillating voltage $\tilde V_G = V \cos(\Omega t)$ is applied on a side-gate electrode which is positioned below the suspended part of the sheet. 

The potential $\varphi$ on the membrane, and by this means the electronic flow through the tunneling barrier,  depends  on the charge $q$ accumulated on the membrane, gate voltage and mutual capacitance. The latter depends on the membrane deflection. At the same time, high frequency electrical field produced by the gate electrode sets up a
time-varying force acting on the charged membrane. This force provides a feedback between the mechanical and the electronic subsystems. We show that such feedback may generate the electromechanical instability resulting in high amplitude mechanical self-oscillations even for the case when the external frequency $\Omega$ is much higher than the characteristic mechanical frequency $\omega_m $.

For a quantitative description of the above-mentioned phenomena, we suggest the following model. We represent the graphene sheet as an elastic thin membrane which motion is described within the continuum mechanics approach and completely characterized by the amplitude of its fundamental bending mode $u(t)$. Since the instability, as it is shown below, is a nonresonant phenomenon, we disregard the geometric nonlinearity of the graphene membrane. The time evolution of the membrane deflection $u$ is described as a damped oscillator subjected to an external electrostatic force induced by the side-gate voltage $\tilde{V}_G$ :
\begin{equation}
\label{fullDyn}
 \ddot u+\frac{\omega_m}{Q} \dot u+\omega_m^2u=\frac{1}{2m}\frac{\partial C_{G}}{\partial u}\left(\varphi-\tilde{V}_G(t)\right)^{2},
\end{equation}
where $Q$ is the quality factor of the oscillator and $C_G (u)$ is the mutual capacitance between the graphene
and the side-gate. Since the membrane deflection $u(t)$ is much smaller than the distance $d$ between the side-gate and the membrane, we set $C_G (u)\approx C_G(1+u(t)/\tilde{d})$ , where $\tilde{d}=C_{G}(0)/C'_{G}(0)\approx d$.

The electrostatic potential on the membrane $\varphi(q,\tilde V_G,u)$ is given by the expression  $\varphi = (C_G (u)\tilde V_G + q(t))/C(u)$ , where $C(u)=C_G(u)+C_E$ is the total capacitance of the membrane and  $C_E$ is the mutual capacitance between the graphene and the grounded electrodes which is independent of the membrane deflection.

The time evolution of the charge $q(t)$ may be described by the following  equation for the equivalent RC-circuit, shown in Figure \ref{fig1}:
\begin{equation} \label{qdynamics}
\dot q  =-\nu_{RC}(u)\left  ( q(t)+C_{G}(u)\tilde{V}_{G} \right)
\end{equation}
where  $\nu_{RC}(u) = 1/RC(u)$  is the charge relaxation frequency, $R$ is the tunnel resistance between the graphene membrane and  leads. The set of (Eq. \ref{fullDyn}) and (Eq. \ref{qdynamics}) describes a coupled dynamics of the electronic and mechanical subsystems.

\begin{figure}\centering
\includegraphics[scale=0.65]{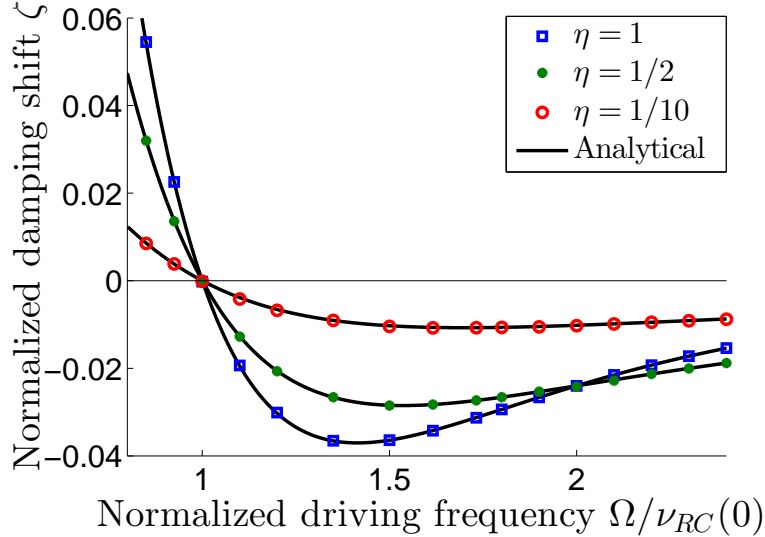}
\caption{Markers indicate numerical simulations of the normalized damping shift of (Eq. \ref{fullDyn}) and (Eq. \ref{qdynamics}) for different values of $\eta$ and parameter values $\omega_m/\nu_{RC}(0)=10^{-2}$ and $u/d\sim 10^{-2}$. The solid black lines are the corresponding analytical solutions $\zeta$ from (Eq. \ref{lambda}). When the driving frequency exceeds the inverse retardation time $\nu_{RC}(0)$ the damping shifts becomes negative. }
\label{fig2}
\end{figure}

To  analyse these equations analytically,
let us introduce  dimensionless variables for the displacement $x= u/\tilde{d}$, time $\tau=t\nu_{RC}(0)$ and charge $\mathrm{q}=q/C_{G}(0)V_{0}$.
Then, by solving (Eq. \ref{qdynamics}) and keeping only linear terms with respect to $x$, we get the following system of equations:

\begin{eqnarray}
\label{uscaled}
 x_{\tau\tau}(\tau)+\frac{\tilde{\omega}_{m}}{Q}x_{\tau} +\tilde{\omega}_{m}^{2}x=\varepsilon \frac{C(0)^{2}}{C(x)^{2}}\left[\eta \mathrm{q}-(1-\eta)\cos(\tilde{\Omega}\tau)\right]^{2}    \\
 \mathrm{q} = -\frac{\cos(\tilde{\Omega}\tau-\vartheta)}{\sqrt{1+\tilde{\Omega}^{2}}}
 -\int_{-\infty}^{\tau}d\tau'e^{\tau-\tau'}\left[\frac{\cos(\tilde{\Omega}\tau'-\vartheta)}{\sqrt{1+\tilde{\Omega}^{2}}}+
 (1-\eta)\cos(\tilde{\Omega}\tau')\right]x(\tau')
\label{qint}
\end{eqnarray}
Here $\tilde{\omega}_{m}=\omega_{m}/\nu_{RC}(0)$, $\tilde{\Omega}=\Omega/\nu_{RC}(0)$,              $\tan\vartheta=\tilde{\Omega}$, $\eta=C_{G}(0)/C(0)$. The parameter $\varepsilon=\Omega^{2}_{V}/\nu_{RC}^{2}(0)=C_{G}V_{0}^{2}/2m\tilde{d}^{2}\nu_{RC}^{2}(0)$ characterizes the electromechanical coupling strength. In there, $\Omega_{V}$ may be estimated as $\Omega_{V}\approx V_{0}\sqrt{\varepsilon_0/2\varrho \tilde d^{3}}$ where $\varepsilon_0$ is the vacuum permittivity and $\varrho$ is the 2D-mass density of the graphene. For the effective distance $\tilde d=10 $nm, one gets $\Omega_{V}\approx 10^{9}\left[ \mathrm{Hz/V}\right]V_{0}$. Therefore,  for the voltage range $V_{0}<10$mV, one can find that the frequency $\Omega_{V}$ is less than 1 MHz that is much smaller than the typical mechanical frequency $\omega_{m}\approx100$MHz.
From (Eq. \ref{qint}) it follows that the instant charge on the membrane exhibits the time-delayed response, with exponential memory decay, to the mechanical displacement of the membrane.  To study the stability of the system we consider the linear regime with respect to $x$ for the small parameter $\varepsilon<<1$ case. Under such conditions the driving field introduces high frequency components of small order $\varepsilon$ to the mechanical vibration  and we can write $x(\tau)=\bar{x}(\tau)(1+\varepsilon\chi(\tau))$, with $\chi(\tau)=\chi(\tau+\pi\tilde{\Omega}^{-1})$ and $\bar x$ changing slowly on the time scale $\tilde\Omega^{-1}$. Substituting (Eq. \ref{qint}) into (Eq. \ref{uscaled}), we get the  equation for the slow component $\bar{x}$:
\begin{equation}\label{u}
 \bar{x}_{\tau\tau}(\tau)+
 \frac{\tilde{\omega}}{Q}\bar{x}_{\tau}(\tau)+(\tilde{\omega}^{2}_{m}+\varepsilon\alpha)\bar{x}(\tau) -\varepsilon\alpha \int_{-\infty}^{0}d\tau'e^{\tau'}\cos(\tau')
 \bar{x}(\tau+\tau')= \frac{\varepsilon\alpha}{2\eta}
 \end{equation}
where $\alpha=\eta(1+(1-\eta)^{2}\tilde\Omega^2)/(1+\tilde\Omega^2)$. We seek the solution of this equation in the form $\bar{x}=\bar{x}_{0}+\delta\bar{x}(\tau)$, where $\bar{x}_{0}=\varepsilon\alpha/2\eta\tilde{\omega}^{2}$ and $\delta\bar{x}(\tau)=\sum_{i}A_{i}\exp\lambda_{i}\tau$  is the general solution of the corresponding  homogeneous equation. The coefficients $\lambda_{i}$ are the solutions to the following equation:
\begin{equation}\label{DE}
  \lambda^{2}+\frac{\tilde{\omega}}{Q}\lambda+(\tilde{\omega}^{2}_{m}+\varepsilon\alpha)=
  \frac{\varepsilon\alpha(1+\lambda)}{(1+\lambda)^{2}+\tilde{\Omega}^{2}}
\end{equation}
To this end, we restrict the analysis on the regime when the external frequency is of the order of the relaxation frequency  but essentially exceeds the mechanical frequency $\tilde{\Omega}\approx 1>> \tilde{\omega}_{m}$ and the electromechanical coupling  is considered to be weak  so $\tilde{\omega}_{m}>>\varepsilon\sim\tilde{\omega}_{m}/Q$.
Under these assumptions, the solution of (Eq. \ref{DE}) in the first order approximation on small parameters  is:
\begin{eqnarray}
\label{lambda}
   \lambda=\tilde{\gamma} \pm i\sqrt{\tilde\omega_{m}^{2}+\epsilon\alpha\frac{\tilde\Omega^2}{1+\tilde\Omega^2}} \nonumber \\
   \tilde{\gamma}=-\frac{1}{2}\left(\frac{\tilde\omega_{m}}{Q}+\epsilon\zeta\right),\ \; \zeta
   =\alpha\frac{1-\tilde{\Omega}^{2}}{(1+\tilde{\Omega}^{2})^{2}}
\end{eqnarray}
From (Eq. \ref{lambda}) one can see that the electromechanical coupling shifts the effective damping of the membrane with respect to the intrinsic one. If the effective damping is negative, the static state $ u(t)=\tilde{d}x_{0}$ is unstable and $\bar x$ performs oscillations with frequency  $\sim\omega_{m}$ and exponentially growing amplitude. This situation is possible when the normalized damping shift $\zeta<0$, i.e., when the external frequency exceeds the charge relaxation frequency $\Omega>\nu_{RC}(0)$. Under this condition, the external high frequency electromagnetic field generates  an effective pumping of the mechanical oscillations despite the fact that the resonance condition is strongly violated $\Omega>>\omega_{m}$. The behavior of the function $\zeta(\tilde{\Omega})$ for different values of $\eta$ is shown in Figure \ref{fig2}. Futher, the instability occurs when the amplitude of the side-gate voltage is larger than a critical value $V_{0}>V_{0}^{c}$ satisfying $\Omega^{2}_V(V_{0}^{c})=-\omega_{m}\nu_{RC}(0)^2/\zeta Q$ so that the pumping overcomes the intrinsic damping. 

To estimate the amplitude of the gate voltage oscillations needed to initiate the instability,  we set $\eta=C_{G}/C=0.5$.
From Figure \ref{fig2} one can deduce that under this condition, the minimum value of $\zeta\sim$-0.03.
Then, by using the following values for the distance $\tilde{d}\approx$100 nm,
 vibrational frequency $\omega_m\approx$ $10^8$Hz, quality factor $Q\approx10^{4}$,  charge relaxation frequency $\nu_{RC}(0)\approx 10^{9}$Hz, membrane tunneling resistance $R\approx10^7 \Omega$, and capacitance of the membrane $C\approx10^{-16}$ F, we get an estimation for the threshold voltage $V_c=\sqrt{2\omega_m\nu_{RC}(0)\rho\tilde d^3(|\zeta| Q \varepsilon_0)^{-1}}\approx10 $mV.
Now, let us analyse the saturation mechanism of the instability. If the system is driven resonantly, the nonlinearity in the mechanical subsystem will saturate the system at a stationary amplitude of oscillations. However, since in our case the  excitation phenomena  is not of the resonance nature, the saturation mechanism, should be of different origin. One of the most possible candidates is the intrinsic nonlinear damping\cite{eimo+11,crmi+12}. It may be taken into account by adding into (Eq. \ref{fullDyn}) the term $\omega_mQ^{-1} \dot u (u-u_{st})^2/a^2$ where $a$ is the characteristic length of damping variations. Such nonlinear damping
saturates the amplitude of the mechanical oscillations at
\begin{equation}
A_{st}=2a\sqrt{\left(\frac{V}{V_c}\right)^2-1}.
\end{equation}
\begin{figure}\centering
\includegraphics[scale=0.8]{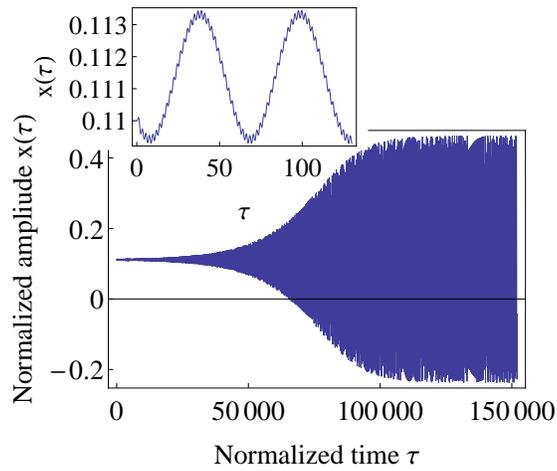}
\caption{Time evolution of the membrane deflection obtained by
direct numerical simulations of (Eq. \ref{fullDyn}) and (Eq. \ref{qdynamics}) including the nonlinear damping in the form discussed in the text. The parameters $\epsilon=10^{-2}/2,\ \Omega/\nu_{RC}(0)=3/2,\ \omega_m/\nu_{RC}(0)=10^{-1},\ Q=10^4$ and $a=d/20$ were chosen in order to clearly demonstrate (a) the exponential amplitude growth and saturation due to the nonlinear damping and (b) small amplitude modulation of fast oscillations with frequency $2\Omega$. This modulation is not visible in (c) since the $\omega_m$ component has been pumped to a relatively large amplitude.}
\label{fig3}
\end{figure}
To confirm the above mentioned phenomena, we perform direct numerical simulations of (Eq. \ref{fullDyn}) and (Eq. \ref{qdynamics}), shown in Figure \ref{fig3}.  

Finally, note that the membrane deflections induced by the ac-voltage gives rise to a nonlinear correction to the impedance $Z_\Omega$ of the equivalent RC-curcuit, see Figure  \ref{fig1}. In particular, one can show that the instability discussed above manifests itself as a discontinuous jump at $V=V_c$ in the first derivative of the average active ac-power W(V)
\begin{equation}
 \frac{\partial W(V)}{\partial V}\bigg|^{V_c^{+0}}_{V_c^{-0}}=I_0(V_c)\beta\left(\frac{a}{d}\right)^2
\end{equation}
where $I_0(V_c)= Re\ Z_c^{-1}V_c$ and $\beta(\eta,\tilde\Omega)$ is a numerical factor. Modeling $C_G$ as a parallel plate capacitor and considering the case $\tilde\Omega=3/2$, which yields almost  optimal pumping regime, we obtain $\beta\approx12(1-0.7\eta)^2$.
  Therefore, measurement of the ac-power may provide information about the electromechanical instability, in particular, to estimate the nonlinear damping of the mechanical vibrations.

To conclude, we have found that the mechanical vibrations of the graphene-based oscillator may be actuated by  nonresonant, high frequency electromagnetic fields. This is due to the fact that the effective damping of the oscillator is reduced when the frequency of the electromagnetic field exceeds the inverse response time of the charge oscillations in the graphene membrane. If the field strength is strong enough to overcome the intrinsic damping, the mechanical vibrations become unstable and saturate due to the nonlinear damping. The phenomena should be detectable with the available experimental techniques not only for the graphene membranes but also for other electromechanical oscillators due to the robustness of the predicted mechanism.

\begin{acknowledgement}
The authors thank the Swedish Research Council for funding our research (VR).

\end{acknowledgement}

\bibliography{article}


\end{document}